\documentclass[preprint,preprintnumbers,amsmath,amssymb]{revtex4}
\usepackage{graphicx}
\usepackage{dcolumn} 
\usepackage{bm} 
\begin{document}
\title{Quantum gravitational optics: Effective Raychaudhuri equation}
\author{ N.~Ahmadi$^{a}$\footnote{
Electronic address:~nahmadi@ut.ac.ir} and M.~Nouri-Zonoz $^{a,b}$ \footnote{
Electronic address:~nouri@theory.ipm.ac.ir}}
\address{$^{a}$ Department of Physics, University of Tehran, North Karegar Ave., Tehran 14395-547, Iran. \\
$^{b}$ Institute for studies in theoretical physics and mathematics, P O Box 19395-5531 Tehran, Iran.}
\begin{abstract}
Vacuum polarization in QED in a background gravitational field induces interactions which {\it effectively}
modify the classical picture of light rays, as the null geodesics of spacetime. These interactions violate 
the strong equivalence principle and affect the propagation of light leading to superluminal photon 
velocities. Taking into account the QED vacuum polarization, we study the propagation of a bundle of rays
in a background gravitational field. To do so we consider the perturbative deformation of Raychaudhuri 
equation through the influence of vacuum polarization on photon propagation. We 
analyze the contribution of the above interactions to the optical scalars namely, shear, vorticity and 
expansion using the Newman-Penrose formalism.
\end{abstract}
\maketitle
\section{Introduction}
Vacuum polarization is an essential ingredient of QED whose contribution leads to astonishingly precise
agreement between predicted and observed values of the electron magnetic moment and Lamb shift.
On the other hand being a quantum field theortetic effect it would be interesting to look for its 
implication in semi-classical gravity as a quantum characteristic of the electromagnetic field coupled to
the gravitational field \cite{0}.
Effects of QED interactions such as vacuum polarization on the photon propagation in a classical 
background gravitational field gives 
rise to a wide range of new phenomena. Simple analysis of null rays in classical general relativity implies that a
curved spacetime,compared to the flat case, could be treated as an optical medium with a refractive 
index \cite{1}. Now adding to that QED effects, in the context of semi-classical gravity, 
leads to interesting phenomena 
such as dispersive effects and polarization dependent propagation of photons. One of the main consequences
is found to be the light cone modification in such a way that the 
QED photons do not propagate along the null geodesics of the background geometry. They propagate instead along 
null geodesics of an effective geometry. In many cases depending on the direction and polarization of photons, 
superluminal propagation becomes possible. Quantum characteristics of photon propagation in a curved 
background opens up a whole new field coined quantum gravitational optics (QGO)\cite{2}, Started with 
the pioneering paper by Drummond and Hathrell who considered the effect of vacuum polarization on photon 
propagation in a curved background and continued with detailed study of the same effect for specific 
spacetimes \cite{3}-\cite{4}. Coupling of the electromagnetic field  with curvature in the QED action 
introduced in \cite{0} violates the strong equivalence principle (SEP) in a mass scale comparable to 
the electron mass $m$. The modification of the underlying spacetime geometry felt by photons due to QED 
interactions has some unexpected results. The Newman-Penrose (NP) formalism has been found to be an 
elegant way of analyzing the results in the SEP violating cases. Quantum gravitational optics attributes  
velocity shift and null cone modification at each point to only a single NP scalar for each of the 
Ricci and Weyl tensors, namely $\Phi_{00}$ and $\Psi_{0}$ respectively \cite{5}. The question is what would be the role of other NP scalars in determining the forms of Ricci and Weyl tensors?
 or what, if any, would be the effect of the gravitational field on 
the other photons traveling with the normal speed of light 
Here we shall study the geometric properties 
of {\it physical} null congruences which are families of physical null rays determined by the effective
metric $G_{\mu\nu}$. These are then distinct from the propertries of the {\it geometric} null congruences characterized by 
the spacetime metric $g_{\mu\nu}$. These congruences are special in the sense that the modification of the light 
cone in a local inertial frame (LIF) can be described as $\left(\eta_{\left(a\right)\left(b\right)}+
\alpha^{2}\sigma_{\left(a\right)\left(b\right)}\left(R\right)\right)k^{\left(a\right)}k^{\left(b\right)}=0$, 
where $\eta_{\left(a\right)\left(b\right)}$ is the Minkowski metric, $\alpha$ is the fine structure 
constant and $\sigma_{\left(a\right)\left(b\right)}\left(R\right)$ depends on the Riemann curvature at 
the origin of the LIF. In one-loop approximation, these photons travel with unit velocity. For on-shell 
photons, some of the NP scalars play the role of optical scalars and their propagation, along the rays, is governed by 
the known Raychaudhuri equation. We show that these scalars must be effectively modified to describe 
the {\it physical} null congruences and find that a new set of NP scalars contribute in the definition 
of these {\it effective optical scalars}. The modified Raychaudhuri equation governing their propagation will be 
discussed. The effective geometry may have features that are seen only by photons when propagating in 
the geometry. There may be hidden singularities in the form of geodesic incompleteness which can be 
studied through the implementation of the effective Raychaudhuri equation.
\section{Quantum gravitational optics}
The effect of one-loop vacuum polarization on the photon propagation in a fixed curved background 
spacetime is represented by the following effective action derived by Drummond and Hathrell \cite{0},
\begin{equation}
\Gamma=\int dx \sqrt{-g}\left[-\frac{1}{4}F_{\mu\nu}F^{\mu\nu}+\frac{1}{m^2}\left(aRF_{\mu\nu}F^{\mu\nu}+
bR_{\mu\nu}F^{\mu\lambda}F^{\nu\ \ }_{\ \ \lambda}+cR_{\mu\lambda\nu\rho}
F^{\mu\lambda}F^{\nu\rho}\right)\right].\label{1}\end{equation} 
Here, $a=-\frac{1}{144}\frac{\alpha}{\pi}$, $b=\frac{13}{360}\frac{\alpha}{\pi}$ and $c=
-\frac{1}{360}\frac{\alpha}{\pi}$ where $\alpha$ is the fine structure constant and $m$ is the electron 
mass. The notable feature in the above action is the direct coupling of the electromagnetic field 
to the curvature tensor 
which in effect violates the strong equivalence principle. When we consider the equations of 
motion derived from this action the Bianchi identity does not change but it gives rise to curvature 
dependent modifications to the equation of motion in the form
\begin{equation}
D_{\mu}F^{\mu\nu}+\frac{1}{m^2}\left(2bR^{\mu \ \ }_{\ \ \lambda}D_{\mu}F^{\lambda\nu}+
4cR^{\mu\nu \ \ }_{ \ \ \lambda\rho}D_{\mu}F^{\lambda\rho}\right)=0
.\label{2}\end{equation}
There are some approximations under which this equation of motion was obtained. The first one is the low 
frequency approximation in the sense that the derivation is only applicable to wavelengths 
$\lambda>\lambda_{c}$. By this approximation we ignore terms in the effective action involving 
higher order field derivatives. The second is a weak field approximation for gravity. 
This implicitly means 
that the wavelengths $\lambda_c << L$ are considered, where $L$ is a typical curvature scale.\\
The characteristics of the light propagation can be studied by applying the geometric optics 
approximation. The electromagnetic field is written in the form $A_{\mu}=Aa_{\mu}e^{i\theta}$, 
where $A$ is the amplitude and $a_{\mu}$ is the polarization vector. The amplitude is taken to be 
slowly-varying in comparision with $\theta$. In each small region of space, we can speak of a 
direction of propagation, normal to a surface at all of whose points the phase of the wave is 
constant. If we now identify the wave vector as $k_{\mu}=\partial_{\mu}\theta$, the equation 
governing the components of this vector field for a photon with spacelike, normalized polarization 
vector i.e, $a_{\mu}a^{\mu}=-1$, can be read from (\ref{2}) as
\begin{equation}
k^2-\frac{2b}{m^2}R_{\mu\lambda}k^{\mu}k^{\lambda}+\frac{8c}{m^2}R_{\mu\nu\lambda\rho}
k^{\mu}k^{\lambda}a^{\nu}a^{\rho}=0.\label{3}\end{equation} 
Eq. (\ref{3}) is an effective light cone equation, re-expressed in terms of the Weyl tensor is given by
\begin{equation}
k^2-\frac{\left(2b+4c\right)}{m^2}R_{\mu\lambda}k^{\mu}k^{\lambda}+\frac{8c}{m^2}C_{\mu\nu\lambda\rho}
k^{\mu}k^{\lambda}a^{\nu}a^{\rho}= 0.\label{4}\end{equation}
The corresponding momentum of photon, $p^{\mu}$, is the tangent vector to the light ray $x^{\mu}$ i.e, 
$p^{\mu}=\frac{d}{ds}x^{\mu}\left(s\right)$. Equation (\ref{4}) being quadratic and homogenous could be 
written in the following form
\begin{equation}
{\cal G}^{\mu\nu}k_{\mu}k_{\nu}=0
\label{4.1}\end{equation}
This equation implies a nontrivial relation between $k_\mu$ and $p^\mu$, through ${\cal G}$ 
by the following argument. 
Refering to the defitions of $k_\mu$ and $p^\mu$, the former defined as the gradient of a phase, constitutes
the component of a one-form and hence belongs to a cotangent space whereas the latter being a tangent vector
belongs to the tangent space. In the usual free electromagnetic theory they are related through the metric 
tensor i.e $k_\mu = g_{\mu\nu}p^\nu$. In the modified theory however, using equations (\ref{4}) and 
(\ref{4.1}), we arrive  at the following non-trivial relation, 
\begin{equation}
k_{\mu}={G}_{\mu\nu}p^{\nu}=p_{\mu}+\frac{1}{m^2}\left[\left(b+2c\right)R_{\mu\lambda}p^{\lambda}
-\left(4c\right)C_{\mu \lambda\sigma\kappa}a^{\lambda}a^{\kappa}p^{\sigma}\right]
\label{5}\end{equation}
where $ G={\cal G}^{-1}$. This shows that at this level of approximation, QGO can be characterized as a 
bimetric theory. The physical light cones are determined by the effective metric ${G}_{\mu\nu}$, and 
are distinct from the geomerical null cones which are fixed by the spacetime metric $g_{\mu\nu}$. 
(Indices are always raised and lowered using the spacetime metric $g_{\mu\nu}$). \\
It should be noted that all these equations are manifestly local Lorentz invariant though the presence 
of the explicit curvature coupling in the effective action means different dynamics in the LIFs at 
different points in spacetime. In this sense these equations violate the strong principle of 
equivalence (SEP). Some of their implications including gravitational birefringence and superluminal speed of light  
 have been discussed in \cite{6}. Equation (\ref{5}) can be employed to examine the SEP violating effects which arise in the physical photon congruence. In the present study we will focus on the modifications produced by the vacuum polarization effects on the propagation of light rays in a curved background.
\section{Effective optical scalars} 
The physical congruence is specified by the integral curves $\gamma$ of the vector field $k^{\mu}$ 
parametrized by the parameter $u$ and scaled such that
${\nabla}_{k}u=1$. In order to examine the relation between the neighboring 
curves of the congruence, the connecting vector field $q^{\mu}$ is introduced \cite{7}.  
Defined along a particular curve $\gamma$, it characterizes the displacement from a point $P\in\gamma$ 
to a 
point $P^{'}$ on a neighboring curve, where $P$ and $P^{'}$ have the same parameter value $u$. 
Mathematically, this means that  $q^{\mu}$ is being Lie transported along the curve by the vector 
field $k^{\mu}$.
\begin{eqnarray}
{\cal{L}}_{k}q^{\mu}=0\ \
{\rm that}\ \ \rm is \ \
{\nabla}_{k}q^{\mu}={\nabla}_{q}k^{\mu}.
\label{6}\end{eqnarray}
 Here the notation ${\nabla}_{X}=X^{\alpha}\nabla_{\alpha}$ for directional covariant derivative is used. 
In a SEP violating theory like QGO, it is illuminating to work in the NP tetrad formalism. Following 
closely the notations used in \cite{6} and \cite{7}, the first step is to choose a null tetrad. We choose $l^{\mu}$ be 
a null vector along the photon momentum. Let $a^{\mu}$ and $b^{\mu}$ be spacelike transverse vectors 
(in the present case they will be identified with the polarization vectors later) and define the complex 
null vectors $m^{\mu}$ and $\overline{m}^{\mu}$ by $m^{\mu}=
\frac{1}{\sqrt{2}}\left(a^{\mu}+ib^{\mu}\right)$ and $\overline{m}^{\mu}=\frac{1}{\sqrt{2}}\left(a^{\mu}-
ib^{\mu}\right)$. Adding to this set another real null vector $n^\mu$, they satisfy the usual NP orthogonality conditions
\begin{equation}
l.m=l.\overline{m}=n.m=n.\overline{m}=0,
\label{7}\end{equation}
and
\begin{equation}
l.l=n.n=m.m=\overline{m}.\overline{m}=0,
\label{8}\end{equation}
The normalization conditions 
\begin{equation}
l.n=-m.\overline{m}=1,
\label{9}\end{equation}
are also imposed. We assign a vierbein $e^{\mu}_{\ \ \left(a\right)}$ to this tetrad as follows
\footnote{Tetrad indices are enclosed in parantheses to be distinguished from the tensor indices.} 
\begin{equation}
e_{\left(1\right)}=e^{\left(2\right)}=l\qquad
e_{\left(2\right)}=e^{\left(1\right)}=n\qquad
e_{\left(3\right)}=-e^{\left(4\right)}=m\qquad
e_{\left(4\right)}=-e^{\left(3\right)}=\overline{m},
\label{10}\end{equation}with the frame metric of the form
\begin{equation}
\eta_{\left(a\right)\left(b\right)}=e^{\mu}_{\ \ \left(a\right)}e_{\mu\left(b\right)}=\left(
\begin{array}{cccc}
0&1&0&0\\
1&0&0&0\\
0&0&0&-1\\
0&0&-1&0
\end{array}\right).\label{11}\end{equation}
Weyl tensor being trace free is completely specified in this basis, by the following five complex scalars \cite{8},
\begin{eqnarray}
\Psi_{0}&=&-C_{abcd}l^{a}m^{b}l^{c}m^{d}=
-C_{1313}\nonumber \\
\Psi_{1}&=&-C_{abcd}l^{a}n^{b}l^{c}m^{d}=
-C_{1213}\nonumber\\
\Psi_{2}&=&-C_{abcd}l^{a}m^{b}\overline{m}^{c}n^{d}=
-C_{1342}\nonumber\\
\Psi_{3}&=&-C_{abcd}l^{a}n^{b}\overline{m}^{c}n^{d}=
-C_{1242}\nonumber\\
\Psi_{4}&=&-C_{abcd}n^{a}\overline{m}^{b}n^{c}
\overline{m}^{d}=-C_{2424}
.\label{12}\end{eqnarray}
The ten components of the Ricci tensor are also 
defined in terms of the following four real and three complex scalars.
\begin{eqnarray}
\Phi_{00}&=&-\frac{1}{2}R_{11};\ \ \ \Phi_{22}=-\frac{1}{2}R_{22};\ \ \ \Phi_{02}=
-\frac{1}{2}R_{33};\ \ \ \Phi_{20}=-\frac{1}{2}R_{44};\ \ \nonumber\\ \Phi_{01}&=
&-\frac{1}{2}R_{13};\ \ \ \Phi_{10}=-\frac{1}{2}R_{14};\ \ \ \Phi_{12}=
-\frac{1}{2}R_{23};\ \  \ \Phi_{21}=-\frac{1}{2}R_{24};\ \ \nonumber\\
\Phi_{11}&=&-\frac{1}{4}\left(R_{12}+R_{34}\right);\ \ \Lambda=\frac{1}{24}R=
\frac{1}{12}\left(R_{12}-R_{34}\right)
.\label{13}\end{eqnarray}
In tetrad formalism, various quantities called "Ricci rotation coefficients", 
$\gamma_{\left(c\right)\left(a\right)\left(b\right)}$, are employed to account for the tetrad covariant 
differentiation. These quantities can be defined as
\begin{equation}
e_{\left(a\right)\mu;\nu}=e_{\ \ \mu}^{\left(c\right)}\gamma_{\left(c\right)\left(a\right)
\left(b\right)}e_{\ \ \nu}^{\left(b\right)}
.\label{14}\end{equation}
In NP formalism, they are called "spin coefficients" and are designated by the following special symbols \cite{8}:
\begin{eqnarray}
\kappa&=&\gamma_{\left(3\right)\left(1\right)\left(1\right)};\ \ \ \rho=\gamma_{\left(3\right)
\left(1\right)\left(4\right)};\ \ \ \epsilon=\frac{1}{2}\left(\gamma_{\left(2\right)\left(1\right)
\left(1\right)}+\gamma_{\left(3\right)\left(4\right)\left(1\right)}\right); \nonumber\\
\sigma&=&\gamma_{\left(3\right)\left(1\right)\left(3\right)};\ \ \ \mu=\gamma_{\left(2\right)
\left(4\right)\left(3\right)};\ \ \ \gamma=\frac{1}{2}\left(\gamma_{\left(2\right)\left(1\right)
\left(2\right)}+\gamma_{\left(3\right)\left(4\right)\left(2\right)}\right); \nonumber\\
\lambda&=&\gamma_{\left(2\right)\left(4\right)\left(4\right)};\ \ \ \tau=\gamma_{\left(3\right)
\left(1\right)\left(2\right)};\ \ \ \alpha=\frac{1}{2}\left(\gamma_{\left(2\right)\left(1\right)
\left(4\right)}+\gamma_{\left(3\right)\left(4\right)\left(4\right)}\right); \nonumber\\
\nu&=&\gamma_{\left(2\right)\left(4\right)\left(2\right)};\ \ \ \pi=\gamma_{\left(2\right)
\left(4\right)\left(1\right)};\ \ \beta=\frac{1}{2}\left(\gamma_{\left(2\right)\left(1\right)
\left(3\right)}+\gamma_{\left(3\right)\left(4\right)\left(3\right)}\right)
.\label{15}\end{eqnarray}
A general rule to obtain the complex conjugate of any quantity is to replace the index (3), wherever 
it occurs with the index (4) and vice versa.\\Using the NP formalism has some advantages, including the fact that 
entirely scalar quantities are used. Also a considerable economy of notation is achieved when complex 
spin coefficients are used instead of the Christoffel symbols employed in the 
conventional coordinate approach. This formalism is most advantageous if the 
null tetrads introduced could be completely tied to the geometry of the problem. Otherwise some freedom remains in 
the choice of the basis as a result of subjecting it to the Lorentz transformation. This has the effect that many of the 
quantities involved in the calculation are 
of the same nature as gauge quantities whose values transform in certain ways as the basis frame is varied in 
accordance with the remaining freedom. In QGO, one null direction is singled out and that is the unperturbed 
photon momentum which is fixed in the direction of the null vector $l^{\mu}$. Therefore the allowed 
transformation would be those which leave the $l^{\mu}$ direction unchanged and preserve the 
underlying orthogonality and normalization conditions. These are classified as follows:
\begin{eqnarray}
{\rm{class }}\ \ I &:&\ \ l\longmapsto l;\ \ m\longmapsto m+al;\ \ \overline{m}\longmapsto \overline{m}+
a^{*}l;\ \ n\longmapsto n+a^{*}m+a\overline{m}+aa^{*}l;\nonumber \\
{\rm{class }}\ \ II &:&\ \ l\longmapsto A^{-1}l;\ \ n\longmapsto An;\ \ m\longmapsto e^{i\theta}m; 
\ \ \overline{m}\longmapsto e^{-i\theta}\overline{m}
,\label{16}\end{eqnarray}
where $a$ is a complex function and $A$ and $\theta$ are two real functions on the manifold.
We shall now proceed to write down  equation (\ref{6}) in the above tetrad basis. Before doing so we employ all the 
gauge freedom we have to fix the tetrad basis. As the tetrad frame 
is transformed, different tetrad components are subject to changes. With $l$ vectors as the velocity 
vector along the geometric congruence, we have
\begin{equation}
{\nabla}_{l}l^{\mu}= \left(\epsilon+\epsilon^{*}\right)l^{\mu}-\kappa \overline{m}^{\mu}-
\kappa^{*}m^{\mu}\ \ \propto l^{\mu}\ \ \Longrightarrow\kappa=0
,\label{17}\end{equation}
and if they are affinely parametrized
\begin{equation}
{\nabla}_{l}l^{\mu}=0\ \ \Longrightarrow\kappa=\epsilon=0.
 \label{18}\end{equation}
If $\kappa=0$, the latter requirement can be met by a class II rotation which will not affect the 
direction of $l$ nor of initially vanishing of $\kappa$. By a suitable rotation of class I, it could be
arranged so that $\pi=0$. So the gauge we are working in is the one in which
$\kappa=\epsilon=\pi=0$. After such a rotation, the newly oriented vectors $n,m$ and $\overline{m}$ will 
remain unchanged as they are parallely propagated along $l$. This could be invoked through the following 
relations
\begin{eqnarray}
{\nabla}_{l}m^{\nu}&=& \pi^{*}l ^{\nu}-\kappa n^{\nu}+ \left(\epsilon-\epsilon^{*}\right)m^{\nu} \nonumber \\
{\nabla}_{l}n^{\nu}&=& \pi^{*}\overline{m}^{\nu}+\pi m^{\nu}- \left(\epsilon+\epsilon^{*}\right)n^{\nu} 
\label{19}\end{eqnarray}
which vanish in the above gauge. Now we have exploited all the gauge freedom we had and 
 ready to write equation (\ref{6}) in the above gauge. We only consider the states which propagate with a 
well defined polarization \footnote{A linearly polarized light may be considered as a superposition of 
a left and a right circularly polarised lights with a phase difference which represents the direction of 
linear polarization. As the light passes through the birefringent medium, the right and the left circularly 
polarised lights propagate with different speeds. So one of the physical manifestation of the birefringence is 
the rotation of the polarisation plane of the linearly polarized light. However, expecting any SEP 
violations to be tiny, we are allowed to neglect the rotation and focus on astronomical sources with a well defined 
polarization.}. For two transverse polarizations expressed in terms of the vectors $m^{\mu}$ and $\overline{m}^{\mu}$, representing the left and the right handed circular polarizations, we have
\begin{equation}
{\nabla}_{k}q^{\nu}={\nabla}_{q}\left[l^{\nu}+\frac{1}{m^2}\left(b+2c\right)R^{\lambda\nu}l_{\lambda}
\mp\frac{2c}{m^2}C^{\nu\ \ \ \ }_{\ \lambda\sigma\kappa}l^{\sigma}\left( m^{\lambda}
\pm\overline{m}^{\lambda}\right)\left(m^{\kappa}\pm\overline{m}^{\kappa}\right)\right]
.\label{20}\end{equation}
After multiplying by $e_{\nu}^{\left(c\right)}$, we get
\begin{equation}
{\nabla}_{k}q^{\left(c\right)}=\left(\eta^{\left(c\right)\left(a\right)}+\frac{1}{m^2}A^{\left(c\right)
\left(a\right)}\right)q^{\left(b\right)}\gamma_{\left(a\right)\left(1\right)\left(b\right)}+
q_{\left(a\right)}\gamma^{\left(a\right)\left(c\right)\left(b\right)}k_{\left(b\right)}
,\label{21}\end{equation}
where 
\begin{eqnarray}
A^{\left(c\right)\left(a\right)}&:=&\left(b+2c\right)R^{\left(c\right)\left(a\right)}
\mp\left(2c\right)M^{\left(c\right)\left(a\right)},\nonumber\\
M^{\left(c\right)\left(a\right)}&:=&C^{\left(c\right)\left(3\right)\left(a\right)\left(3\right)}
+ C^{\left(c\right)\left(4\right)\left(a\right)\left(4\right)} \pm\left(C^{\left(c\right)\left(3\right)
\left(a\right)\left(4\right)}+C^{\left(c\right)\left(4\right)\left(a\right)\left(3\right)}\right)
.\label{22}\end{eqnarray}
The specifically gravitational birefringence shows up here in $M^{\left(a\right)\left(c\right)}$. 
(we note that $A^{\left(c\right)\left(a\right)}=A^{\left(a\right)\left(c\right)}$). 
Equation (\ref{21}) is a system of coupled differential equations describing the propagation of the tetrad components 
of the connecting vector along the velocity vector $k^{\mu}$. The second term in (\ref{21}) appears as a 
consequence of the null tetrad propagation along the {\it physical} ray (with the tangent vector 
$k^\mu$) i.e $q^\nu {\nabla}_{k} e_{\nu}^{\left(c\right)}$.
We introduce different tetrad components of $q^{\mu}$ 
in the following way
\begin{equation}
q^{\mu}=gl^{\mu}+\xi\overline{m}^{\mu}+\overline{\xi}m^{\mu}+hn^{\mu}
.\label{23} \end{equation}
so that 
\begin{equation}
q^{\left(1\right)}=q_{\left(2\right)}=g;\ \ q^{\left(2\right)}=q_{\left(1\right)}=h;\ \ q^{\left(3\right)}
=-q_{\left(4\right)}=\overline{\xi};\ \ q^{\left(4\right)}=-q_{\left(3\right)}=\xi
.\label{24}\end{equation}
Explicitly for $c=4$, (\ref{21}) gives
\begin{eqnarray}
{\nabla}_{k}q^{\left(4\right)}&=& \left(\eta^{\left(4\right)\left(3\right)}+
\frac{1}{m^2}A^{\left(4\right)\left(3\right)}\right)\left[q^{\left(2\right)}
\gamma_{\left(3\right)\left(1\right)\left(2\right)}+q^{\left(3\right)}
\gamma_{\left(3\right)\left(1\right)\left(3\right)}+q^{\left(4\right)}
\gamma_{\left(3\right)\left(1\right)\left(4\right)}\right]\nonumber\\
&+&\frac{1}{m^2}A^{\left(4\right)\left(4\right)}\left[q^{\left(2\right)}
\gamma_{\left(4\right)\left(1\right)\left(2\right)}+q^{\left(3\right)}
\gamma_{\left(4\right)\left(1\right)\left(3\right)}+q^{\left(4\right)}
\gamma_{\left(4\right)\left(1\right)\left(4\right)}\right]\nonumber\\
&+&\frac{1}{m^2}A^{\left(4\right)\left(2\right)}q^{\left(b\right)}\gamma_{
\left(2\right)\left(1\right)\left(b\right)}-\frac{1}{m^2}A^{\left(2\right)
\left(b\right)}q^{\left(a\right)}\gamma_{\left(a\right)\left(3\right)\left(b\right)}
\label{25}\end{eqnarray}
in which the last term is again the contribution due to the null tetrad propagation along the physical ray discussed
below equation (\ref{22}).
To identify the geometrical and physical content of this equation, a "covariant approach" in which any complicated 
gauge behavior is completely avoided must be chosen. For example, in (\ref{25}),
the term $A^{\left(4\right)\left(4\right)}q^{\left(2\right)}
\gamma_{\left(4\right)\left(1\right)\left(2\right)}$ having such a complicated gauge behavior, transforms into
\begin{equation}
\left[ A^{\left(4\right)\left(4\right)} - 2aA^{\left(4\right)\left(2\right)} + a^2 A^{\left(2\right)\left(2\right)}\right]q^{\left(2\right)}[\gamma_{\left(4\right)\left(1\right)\left(2\right)} + a^*\gamma_{\left(4\right)\left(1\right)\left(3\right)} +
a\gamma_{\left(4\right)\left(1\right)\left(4\right)}]
\label{250}
\end{equation}
under a class $I$ transformation. 
To keep the formalism covariant we remove the gauge-dependent
terms by applying the extra condition that the neighboring pair of rays to satisfy the following condition,
\begin{equation}
q.k=0,  
\label{26}\end{equation} 
and call them {\it abreast} \cite{7}. This is a {\it primary} constraint and it means
\begin{eqnarray}
O\left(\alpha^{0}\right)&:&\ \ q.l=0\Rightarrow h=0;\nonumber \\
O\left(\alpha^{1}\right)&:&\ \ A_{\left(1\right)\left(a\neq 2\right)}=0 \Rightarrow\Phi_{00}=\Phi_{01}
=\Phi_{10}=\Psi_{0}=\Psi_{1}=0.
\label{27}\end{eqnarray}
Eq. (\ref{27}) expresses the abreastness constraint which is independent of the 
parametrization of $k^{\mu}$ as could be easily checked by making the substitution $q^{\mu}\mapsto 
q^{\mu}+\Lambda k^{\mu}$. It is easy to see that under the new parametrization, $q.k=O
\left(\alpha^{2}\right)$, and therefore could be ignored. Furthermore, like any other 
constrained system, we need to make sure that the eq. (\ref{26}) is conserved under the propagation along $k$. 
In other words, ${\nabla}_{k}\left(q.k\right)=0$ does not lead to a {\it secondary} constraint. 
Up to the first order in $\alpha$, we have
\begin{eqnarray}
{\nabla}_{k}\left(q^{\mu}k_{\mu}\right)=0 &\Longrightarrow & \left({\nabla}_{q}k^{\mu}\right)
k_{\mu}+q^{\mu}\left({\nabla}_{k}k_{\mu}\right)=0\nonumber\\
O\left(\alpha^{0}\right)&:&\ \ \left({\nabla}_{q}l^{\mu}\right)l_{\mu}+
q^{\mu}\left({\nabla}_{l}l_{\mu}\right)=0\nonumber\\
O\left(\alpha^{1}\right)&:&\ \ \left[A^{\left(2\right)\left(a\right)}q^{\left(b\right)}+
A^{\left(2\right)\left(b\right)}q^{\left(a\right)}\right]\gamma_{\left(a\right)\left(1\right)\left(b\right)}
=0
,\label{28}\end{eqnarray}
which leads to the same  constraint mentioned above. We emphasize that provided  
the abreastness constraint (\ref{27}) is satisfied, the tangent vector $k^{\mu}$ can be described as 
\begin{eqnarray}
&&k_{\left(2\right)}=1-\frac{1}{m^2}A_{\left(1\right)\left(2\right)},\ \ \ \ k_{\left(a\neq2\right)}
=0,\nonumber \\
&&k^{2}=O\left(\alpha^{2}\right),\ \ \ \ {\nabla}_{k}k^{\mu}=O\left(\alpha^{2}\right)
.\label{29}\end{eqnarray}
The abreast rays are {\it special} physical rays whose effective light cone condition, in 
a local inertial frame, is given by 
\begin{eqnarray}
\left(\eta_{\left(a\right)\left(b\right)}+\alpha^{2}
\sigma_{\left(a\right)\left(b\right)}\left(R\right)\right)k^{\left(a\right)}k^{\left(b\right)}=0.
\label{29-1}\end{eqnarray}
We recall from (\ref{3}) that the general light cone modification induced by the effective action 
(\ref{1}) is of order $\alpha$.\\
For abreast rays, the covariant deformation of $\xi$ implies
\begin{equation}
 {\nabla}_{k}\xi =-\left(\overline{\xi}\sigma+\xi\rho\right) +\frac{1}{m^2}A^{\left(4\right)\left(3\right)}
\left(\overline{\xi}\sigma+\xi\rho\right)+\frac{1}{m^2}A^{\left(4\right)\left(4\right)}\left(\overline{\xi}
\rho^{*}+\xi\sigma^{*}\right)
 .\label{30}\end{equation}
Due to their gauge-dependence and also for our purpose, propagation of the other components of the connecting 
vector $q$ are of less geometric and physical importance and we just state the results below for the sake of 
the completeness of the discussion, 
 \begin{eqnarray}
{\nabla}_{k}h&=&0 \nonumber \\
{\nabla}_{k}g&=&\overline{\xi}\left(\beta+\alpha^{*}\right)+\xi\left(\alpha+\beta^{*}\right)+
\frac{1}{m^2}\left[A^{\left(1\right)\left(3\right)}\left(\overline{\xi}\rho+\xi\sigma^{*}\right)+
c.c\right]
.\label{31}\end{eqnarray}
Now having the equation (\ref{30}) we are ready to study the evolution of the cross sections
of (physical) bundles of rays, formed by their intersection with the spatial 
plane spanned by $m^{\mu}$ and $\overline{m}^{\mu}$. To do so we employ the so called {\it optical scalars}
associated with a bundle of rays, namely the {\it expansion}, the {\it shear} and the {\it twist} \cite{80}. Since the transverse distances from a specified ray $\gamma$ are (frame)observer-independent, for abreast rays these scalars bear explicit physical interpretations.
To simplify the case we consider the propagation along $k$ of the area of a small triangle formed by the 
points $o,\xi_{1}$ and $\xi_{2}$ (see figure 1), we have
\begin{figure}
 \centering
 \includegraphics[angle =0,scale=0.7]{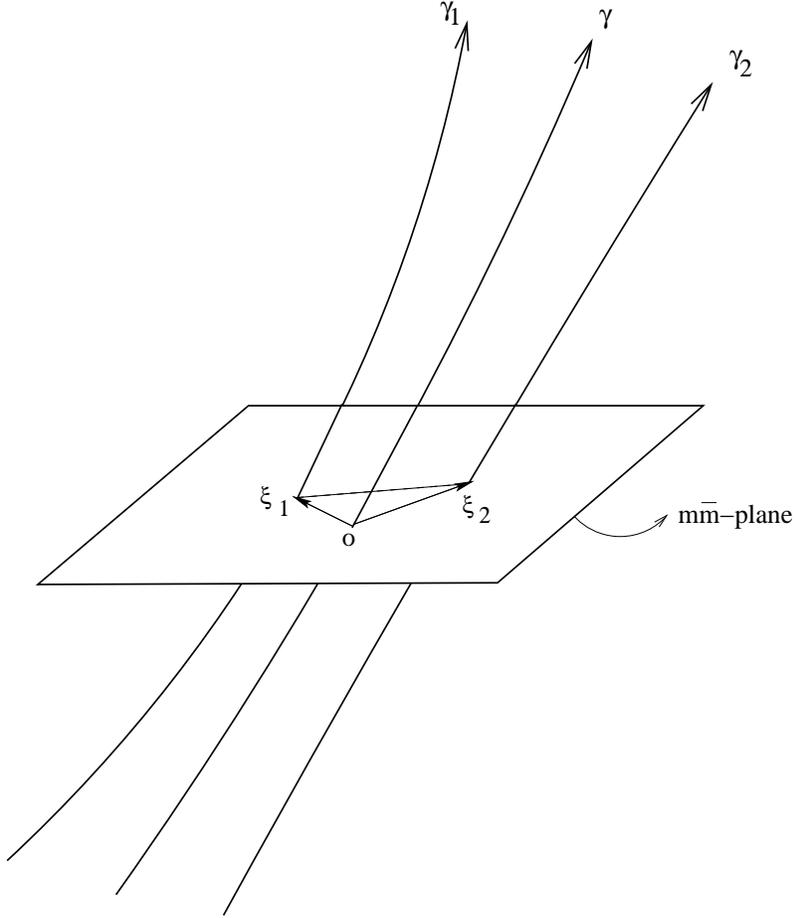}
 \label{fig6.eps}\caption{Three abreast rays of a congruence and the small triangle formed by two connecting vectors $\xi_1$ and  $\xi_2$, connecting the ray $\gamma$
to the rays $\gamma_1$ and $\gamma_2$ in the $m{\bar m}$-plane.}
\end{figure}
\begin{equation}
{\nabla}_{k}\delta A={\nabla}_{k}\left[\frac{i}{2}\left(\xi_{1}\overline{\xi}_{2}-\xi_{2}
\overline{\xi}_{1}\right)\right]=
-\left[\rho -\frac{1}{m^2}\left(A^{\left(3\right)\left(4\right)}\rho+A^{\left(3\right)\left(3\right)}
\sigma\right)+c.c\right]\delta A
\label{32}\end{equation}
The scalar quantity
\begin{equation} -{\rm{Re}}\left[\rho -\frac{1}{m^2}\left(A^{\left(3\right)\left(4\right)}\rho+
A^{\left(3\right)\left(3\right)}\sigma\right)\right]
\label{33}\end{equation}
 is called the expansion parameter, $\theta_{\rm{eff}}$, whose role as a measure of the pattern convergence 
(or divergence) is clear from (\ref{32}). This equation shows that in an effective theory, unlike in the classical case, the local effect of 
$\sigma$ can also change the area. Another useful parameter, the 
"luminosity parameter" $L$, 
is  defined along a bundle of rays as $L^{2}\propto \delta A$ and is related to $\theta_{\rm{eff}}$ 
 as follows (using (\ref{32})), 
\begin{equation}
{\nabla}_{k}L =\theta_{\rm{eff}}L. 
\label{33a}\end{equation}
In other words, 
the expansion parameter is the logarithmic derivative of the luminosity parameter. An alternative 
formula for $\theta_{\rm{eff}}$ is given by $\frac{1}{2}k^{\mu}_{\ \ ;\mu}$.\\
Setting $\theta_{\rm{eff}}$ and the coefficient of $\overline{\xi}$ equal to 
zero in (\ref{30}), we obtain
\begin{equation}
{\nabla}_{k}\xi =-i{\rm{Im}}\left[\rho - \frac{1}{m^2}\left(A^{\left(3\right)\left(4\right)}\rho+A^{\left(3\right)
\left(3\right)}\sigma\right)\right]\xi
,\label{34}\end{equation}
So it could also be claimed that
${\rm{Im}}\left[\rho - \frac{1}{m^2}\left(A^{\left(3\right)\left(4\right)}\rho+A^{\left(3\right)
\left(3\right)}\sigma\right)\right]$
measures the twist in the bundle's cross section and could be called the effective twist, $\omega_{\rm{eff}}$. The fact that this 
combination of scalars is a measure of twist is consistent with the requirement that it measures the failure of $k_\mu$ to be hypersurface orthogonal. In other words
it should vanish for a congruence corresponding to the gradient of a scalar field
i.e,
\begin{equation}
k_{\mu}=\nabla_{\mu}f\Longrightarrow \omega_{\rm{eff}}=0 
\label{35}\end{equation}
To show this, we note that the left hand side of (\ref{35}) is equivalent to 
$\nabla_{[\mu}k_{\nu]}=0$.
At the leading order, $O\left(\alpha^{0}\right)$, this gives $\gamma_{\left(a\right)
\left(1\right)\left(b\right)}=\gamma_{\left(b\right)\left(1\right)\left(a\right)}$,
which in turn, using (\ref{15}), leads to ${\rm{Im}}\rho =0$ for $a=3$ and $b=4$.
The subleading term, $O\left(\alpha^{1}\right)$, is given by
\begin{equation}  
\left[\left(b+2c\right)R_{\lambda\nu}-\left(4c\right)C_{\nu\rho\lambda\sigma}a^{\rho}a^{\sigma}\right]e^{\left(a\right)\lambda}\gamma_{\left(a\right)\left(1\right)
\left(b\right)}e^{\left(b\right)}_{\mu}
=\left[\left(b+2c\right)R_{\lambda\mu}-
\left(4c\right)C_{\mu\rho\lambda\sigma}a^{\rho}a^{\sigma}\right]e^{\left(e\right)\lambda}\gamma_{\left(e\right)\left(1\right)\left(f\right)}e^{\left(f\right)}_{\nu}
\label{36}\end{equation}
Multiplying both sides of (\ref{36}) by $e^{\mu\left(d\right)}e^{\nu\left(c\right)}
e^{\rho\left(g\right)}e^{\sigma\left(h\right)}$, we end up with
\begin{equation}
A^{\left(a\right)\left(c\right)}\gamma_{\left(a\right)\left(1\right)\left(b\right)}
\eta^{\left(b\right)\left(d\right)}=A^{\left(e\right)\left(d\right)}\gamma_{\left(e\right)
\left(1\right)\left(f\right)}\eta^{\left(f\right)\left(c\right)}
\label{37}\end{equation}
Using the abreastness conditions (\ref{27}), for $c=4$ and $d=3$, we obtain 
${\rm{Im}}\left[\rho - \frac{1}{m^2}\left(A^{\left(3\right)\left(4\right)}\rho+A^{\left(3\right)
\left(3\right)}\sigma\right)\right]=0$
, proving our assretion that $\omega_{\rm{eff}}=0$ to the right order.\\
If $\theta_{\rm{eff}}=\omega_{\rm{eff}}=0$, the remaining part of eq. (\ref{30}), i.e
\begin{equation}
\sigma+\frac{1}{m^2}\left(A^{\left(4\right)\left(3\right)}\sigma+A^{\left(4\right)\left(4\right)}
\rho^{*}\right)
,\label{38}\end{equation}
can be interpreted as the effective shear, $\sigma_{\rm{eff}}$, which is a measure of the distortion in the  
 shape of the bundle's cross section so that a circular cross section transforms into an elliptic one. Setting $\sigma_{\rm{eff}}=se^{2i\theta}$, (\ref{30}) reads ${\nabla}_{k}\xi=-se^{2i\theta }\overline{\xi}$. 
This shows that,  $ {\nabla}_{k}\xi$ is a real multiple of $\xi$ when arg$\xi=
\theta, \theta+\pi$ or $\theta\pm\frac{1}{2}\pi$, where for $s>0$, in the first two cases we get contraction towards 
the origin while in the other two cases it experiences dilation. Thus $s$ is a measure of the pattern's shear while $\theta$ 
defines the angle that its minor axis makes with the $\xi$-plane. The quantities 
$\omega_{\rm{eff}}$ and $\sigma_{\rm{eff}}$ can also be defined as
\begin{eqnarray}
 \omega_{\rm{eff}}^{2}&=&\frac{1}{2}k_{[i;j]}k^{i;j}\nonumber\\
 \left|\sigma_{\rm{eff}}\right|^{2}+\theta_{\rm{eff}}^{2}&=&\frac{1}{2}k_{(i;j)}k^{i;j}
 .\label{39}\end{eqnarray} 
\section{Physical null congruence and space time curvature: effective Raychaudhuri equation}
Since the evolution of the cross section of a bundle of rays is characterized by the quantities $\rho$ and $\sigma$ in the one hand and on the other hand the spacetime curvature affects the geometry of null congruences, we need to study the variation of optical scalars along the physical ray . This can be achieved by examining  the second derivative of the connecting vector, $q^{\mu}$, propagating along the wave
vector $k_{\mu}$ through the operation of ${\nabla}_{k}$ on eq. (\ref{1}) ,
\begin{equation}
{\nabla}_{k}{\nabla}_{k}q^{\mu}={\nabla}_{k}\left({\nabla}_{q}k^{\mu}\right)={\nabla}_{q}{\nabla}_{k}k^{\mu}
+R^{\ \ \ \ \mu}_{\lambda\nu\sigma }k^{\lambda}q^{\nu}k^{\sigma}\
= R^{\ \ \ \ \mu}_{\lambda\nu\sigma}k^{\lambda}q^{\nu}k^{\sigma}+O\left(\alpha^{2}\right)
.\label{40}\end{equation}
After multiplying by $e^{\left(c\right)}_{\mu}$ and taking the components corresponding to the 
parallely propagated tetrads we get
\begin{equation}
{\nabla}_{k}{\nabla}_{k}q^{\left(c\right)}=R_{\left(1\right)\left(3\right)\left(1\right)}^{\ \ \ \ \ 
\ \ \  \left(c\right)}\overline{\xi}+R_{\left(1\right)\left(4\right)\left(1\right)}^{\ \ \ \ \ \ \ \  
\left(c\right)}\xi+O\left(\alpha^{2}\right)
.\label{41}\end{equation}
Here, we have neglected the terms including the product of curvature components which would be 
suppressed by $O(\frac{R}{m^2})$. From (\ref{30}), we obtain the useful $\left(2\times2\right)$ 
matrix form as
\begin{equation}
{\nabla}_{k}\mathbf{Z}=\mathbf{P}\mathbf{Z}
,\label{42}\end{equation}
where
\begin{equation}
\mathbf{Z}=\left(
\begin{array}{c}
\xi\\
\overline{\xi}
\end{array}
\right),\ \  
\mathbf{P}=\left(\begin{array}{cc}
\theta_{\rm{eff}}-i\omega_{\rm{eff}}&\ \ \sigma_{\rm{eff}}\\
\sigma^{*}_{\rm{eff}}&\ \ \theta^{*}_{\rm{eff}}+i\omega^{*}_{\rm{eff}}
\end{array}\right)
.\label{43}\end{equation}
For abreast rays (\ref{28}), the equation (\ref{40}) gives
\begin{equation}
{\nabla}_{k}{\nabla}_{k}\mathbf{Z}=\mathbf{0}
.\label{44}\end{equation}
After differentiating (\ref{42}) once more and using (\ref{44}), we get
\begin{equation}
{\nabla}_{k}{\nabla}_{k}\mathbf{Z}=\left({\nabla}_{k}\mathbf{P}\right)\mathbf{Z}+\mathbf{P}
\left({\nabla}_{k}\mathbf{Z}\right)=\left({\nabla}_{k}\mathbf{P}+\mathbf{P}^{2}\right)\mathbf{Z}=
\mathbf{0}.
\label{45}\end{equation}
Since this holds for arbitrary $\xi$, we must have
 \begin{equation}
{\nabla}_{k}\mathbf{P}=-\mathbf{P}^{2}.
\label{46}\end{equation}
Written out in full, eq. (\ref{46}) gives the directional propagation of $\rho$ and $\sigma$ along the physical ray as follows, 

\begin{eqnarray}
{\nabla}_{k}\rho &=&\rho^{2}+\left|\sigma\right|^{2}+\frac{1}{m^2}\left[A^{\left(3\right)\left(4\right)}
\left(\rho^{2}+\left|\sigma\right|^{2}\right)+A^{\left(3\right)\left(3\right)}\rho\sigma+A^{\left(4\right)
\left(4\right)}\rho\sigma^{*}\right]\nonumber\\
{\nabla}_{k}\sigma
&=&\left(\rho+\rho^{*}\right)\sigma+\frac{1}{m^2}\left[A^{\left(3\right)\left(4\right)}\left(\rho+\rho^{*}
\right)\sigma+A^{\left(3\right)\left(3\right)}\sigma^{2}+A^{\left(4\right)\left(4\right)}
\left|\rho\right|^{2}\right]
\label{47}\end{eqnarray}
Compared with the results in the classical case, the above equations could be called
{\it effective Sachs equations}.
With these equations in hand, we can derive the equations governing the directional propagation of the effective optical scalars as
\begin{eqnarray}
{\nabla}_{k}\sigma_{\rm{eff}}&=&-2\theta_{\rm{eff}}\sigma_{\rm{eff}},\\
{\nabla}_{k}\omega_{\rm{eff}}&=&-2\theta_{\rm{eff}}\omega_{\rm{eff}},\label{49}\\
{\nabla}_{k}\theta_{\rm{eff}}&=&\omega_{\rm{eff}}^{2}-\theta_{\rm{eff}}^{2}-\left|\sigma_{\rm{eff}}
\right|^{2}.
\label{50}\end{eqnarray}
These equations are the main results of the paper and their physical importance (consequences) is discussed below.
\section{discussion}
Here we employed the Newmann-Penrose tetrad formalism to derive the  {\it effective Raychaudhuri equation}, i.e the Raychaudhuri equation modified by the vacuum polarization effects on the propagation of a bundle of rays. In the derivation of the classical version of the Raychadhuri equation \cite{9}-\cite{10} the variations of the tangent vector
$k_{\mu;\nu}$ is projected \footnote{The projection operator 
is $P_{\mu\nu}=m_{\nu}\overline{m}_{\mu}+\overline{m}_{\nu}m_{\mu}=g_{\mu\nu}+l_{\mu}n_{\nu}
+n_{\mu}l_{\nu}.$} on the plane spanned by $m^{\mu}$ and $\overline{m}^{\mu}$ and then its trace , 
antisymmetric and symmetric parts are designated as expansion, vorticity and shear
respectively. Here in the modified version we chose, through the equation (\ref{6}), an 
equivalent method in which the variations of the connecting vector $q^\mu$ instead of the tangent vector have been considered and then defined the corresponding {\it effective} optical scalars . The invariants of the theory are easily seen in 
this way. For a luminosity parameter, L, we see from (\ref{49}) that ${\nabla}_{k}\left(L^{2}\omega_{eff}\right)
=0$, in other words here $q.k$ and $L^{2}\omega_{eff}$ are the constant geometric quantities along the congruence.  For two neighbouring rays of $\gamma$, whose connecting vectors $q^{\mu}$ and $\tilde{q}^{\mu}$ 
independently satisfy (\ref{40}), a symplectic invariant is attributed as follows
\begin{equation}
q^{\mu}{\nabla}_{k}\tilde{q}_{\mu}-\tilde{q}^{\mu}{\nabla}_{k}q_{\mu}
.\label{51}\end{equation}
These invariants reduce to the classical ones in the limit of zero perturbation and therefore there remains no anomaly in QGO. 
This is the case since we have applied the symmetries present at $O\left(\alpha^{0}\right)$ as a set of constraints and exploited them to find the covariant quantum corrections. Since the effective action we started with has been derived in a gauge invariant manner \cite{0}, the results obtained are also gauge invaraiant.\\  
The classical limit of the eq (\ref{50}) can be compared with that corresponding to the standard (pure general relativistic) Raychaudhuri equation, namely
\begin{equation} 
{\nabla}_{k}\theta=\omega^{2}-\theta^{2}-\left|\sigma\right|^{2}-\Phi_{00}
.\label{52}\end{equation}
 The last term, i.e. $\Phi_{00}=\frac{1}{2}R_{\left(1\right)\left(1\right)}$, is negative for all 
known forms of ponderable matter. Then for an initially contracting congruence, in the absence of 
rotation and shear ($\rho$ real and $\sigma=0$), the eq. (\ref{52}) results in the divergence of 
the expansion parameter ($\theta\rightarrow-\infty$). This is a necessary though not a sufficient 
condition in singularity theorems \cite{11}. In our case a real $\rho$ and vanishing $\sigma$ is equivalent to $\omega_{\rm{eff}}
=\sigma_{\rm{eff}}=0$. Using (\ref{47}), we see that during the propagation along the ray, they differ 
from zero only by $O\left(\alpha^{2}\right)$,
\begin{eqnarray}
{\nabla}_{k}\theta&=&{\nabla}_{k}\theta_{\rm{eff}}-\frac{1}{m^2}{\nabla}_{l}\left(A_{\left(3\right)
\left(4\right)}\rho\right)\nonumber\\
&=&-\theta^{2}+\frac{1}{m^2}A_{\left(3\right)\left(4\right)}\theta^{2}+O\left(\alpha^{2}\right)
\label{53}\end{eqnarray}
On the other hand, the last term in (\ref{52}) does not appear in our 
calculations due to the abreastness conditions. However, even if $R_{\left(1\right)\left(1\right)}=0$, 
the correction terms calculated in this paper will contribute and their sign must be taken into account.
In QGO, the light cone condition as well as the velocity shift depend on 
$\Phi_{00}$ and $\Psi_{0}$ ( or $A_{\left(1\right)\left(1\right)}$ in our notation), while in the propagation of the expansion 
parameter these are  $\Phi_{11}$, $\Lambda$ and $\Psi_{2}$ (or $A_{\left(3\right)\left(4\right)}$ in our notation) that play role purturbatively. These may have consequences including existence of singularities in the effective geometry that are seen only by QGO photons. The existence of singularities seen by nonlinear photons has been studied in \cite{12}.
The generalization of Drummond-Hathrell results to massless neutrino propagation in a background 
spacetime \cite{13} can be employed to find the corresponding effective Raychaudhuri equation. On the other hand, the effects originated from QED interactions  discussed in this paper, in which local invariance and weak equivalence principle are preserved, are somewhat similar to the phenomenological theories incorporating Lorentz and CPT violating interactions. For example, in a theory describing the propagation of light discussed in \cite{14}, the Lorentz violating coupling constant $K_{\mu\nu\lambda\rho}$ plays the role of the Riemann tensor $R_{\mu\nu\lambda\rho}$ in eq. (\ref{1}). It is therefore easy to translate the phenomenological aspects of the effective Raychaudhuri equation discussed here to this class of theories. We recall that the relative shift in velocity or optical scalars  would be of $O\left(\alpha\frac{\lambda_{c}^2}{L^2}\right)$ for a typical curvature scale $L$.
\section *{Acknowledgments} 
The authors would like to thank University of Tehran for supporting this project under the grants 
provided by the research council.

\end{document}